THEORY

# Advance Sharing of Quantum Shares for Classical Secrets

**RINA MIYAJIMA [1] AND RYUTAROH MATSUMOTO [1,2], (Member, IEEE)**
[1]Department of Information and Communications Engineering, Tokyo Institute of Technology, Tokyo 152-8550, Japan
[2]Department of Mathematical Sciences, Aalborg University, 9220 Aalborg, Denmark

Corresponding authors: Rina Miyajima (miyajima.r.ab@m.titech.ac.jp) and Ryutaroh Matsumoto (ryutaroh@ict.e.titech.ac.jp)

**ABSTRACT** Secret sharing schemes for classical secrets can be classified into classical secret sharing schemes and quantum secret sharing schemes. Classical secret sharing has been known to be able to distribute some shares before a given secret. On the other hand, quantum mechanics extends the capabilities of secret sharing beyond those of classical secret sharing. We propose quantum secret sharing with the capabilities in designing of access structures more flexibly and realizing higher efficiency beyond those of classical secret sharing, that can distribute some shares before a given secret.

## I. INTRODUCTION

Secret sharing scheme [1] is a cryptographic scheme to encode a secret into multiple pieces of information (called shares) and distribute shares to participants so that qualified sets of participants can reconstruct the secret but forbidden sets can gain no information about the secret. For instance, it can be used to guarantee that no individual can obtain an industrial secret, or can launch a nuclear missile, but qualified groups can. The set of qualified sets and that of forbidden sets are called an access structure [2]. In common uses of secret sharing schemes, it is assumed that a dealer can communicate with participants after the dealer obtains a secret.

We consider the following problem: In a country, the president suffers from a serious disease and is anxious about his sudden death. He is afraid that his death makes a national secret accessible to no one if he alone knows about the national secret. For this reason, the president wishes to share the national secret to the dignitaries by a secret sharing scheme. The national secret is sensitive information and the president needs to hand encoded information of the national secret to the dignitaries. The president will obtain the national secret three days later but some dignitaries will make an extended business trip to foreign country from tomorrow. How can the president share the secret?

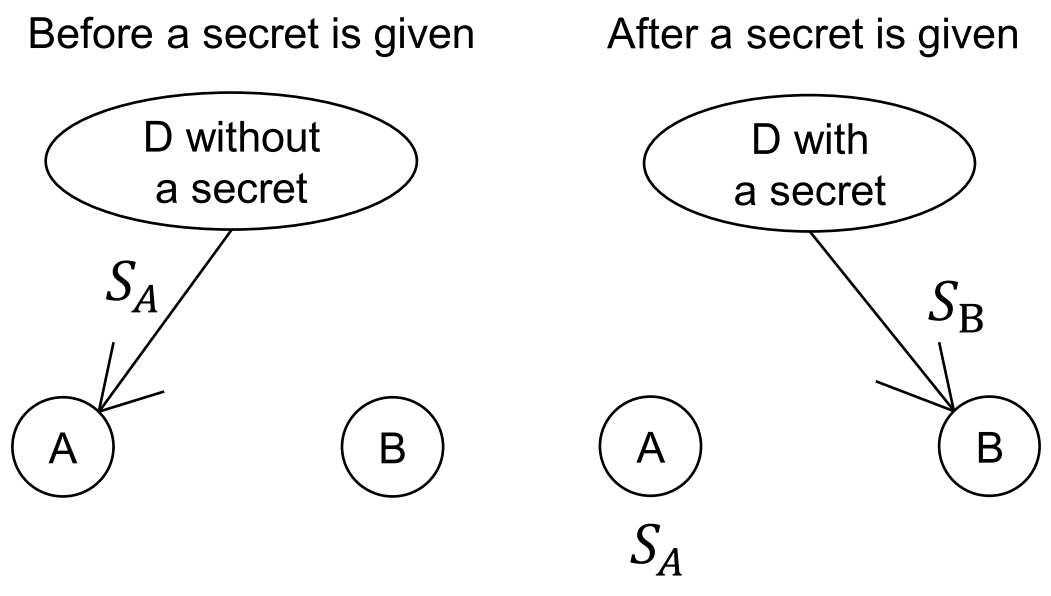

**FIGURE 1. Advance sharing is distribution of shares to some participants before a given secret.**

As we can see from this problem, perhaps a dealer may be unable to communicate with some participants after the dealer obtains a secret. In those situations, it is desirable for the dealer to distribute shares to some participants while the dealer can communicate with participants. To realize this distribution, the dealer needs to be capable to distribute shares to some participants before a given secret. We call this distribution "advance sharing" and a set of shares that can be distributed in advance is called "advance-shareable". For example, a dealer considers to share a 1-bit secret $M$ to participants $A$ and $B$ (see Fig. 1). Before a secret is given, the dealer randomly chooses either 0 or 1 as a 1-bit share $S_A$ and distributes a 1-bit share $S_A$. After the dealer obtains a secret $M$, the dealer generates a share $S_B$ by exclusive-OR ⊕ (XOR), $S_B = M \oplus S_A$ and then distributes a 1-bit share $S_B$ to participant $B$. Only if participants $A$ and $B$ collaborate, they can reconstruct a secret $M$ by XOR $\oplus$, $M = S_A \oplus S_B$. In this example, the share $S_A$ is advance-shareable.

Secret sharing schemes for classical secrets can be classified into classical secret sharing schemes [1], [2], [3], [4] and quantum secret sharing schemes [5], [6], [7], [8], [9], [10], [11], [12]. Classical secret sharing uses classical information as shares while quantum secret sharing uses quantum information [13] as shares. Classical secret sharing has been known to be able to distribute some shares before a given secret. Advance sharing for quantum secrets was proposed in [14] and it can be used for advance sharing for classical secrets. However, an advantage over classical secret sharing has not been clarified in advance sharing of quantum shares for classical secrets by [14]. We propose quantum secret sharing with the capabilities in designing of access structures more flexibly and realizing higher efficiency [2, Definition 13.4] (i.e., the ratio of the size of a secret over the size of each share) beyond those of classical secret sharing, that can distribute some shares before a given secret.

On the other hand, quantum mechanics provides the prominent capabilities to information processing beyond those of classical information processing [13]. For example, quantum computation [15], [16] and quantum teleportation [17], [18], [19] are studied by many researchers. As another application, quantum mechanics extends the capabilities of secret sharing beyond those of classical secret sharing as below [5], [6], [7], [8], [9], [10], [11], [12]. For a fixed size of classical secrets and a fixed size of shares, quantum secret sharing enables designing of access structures more flexibly than classical secret sharing. For example, consider a scheme to share a 2-bit classical secret to 2 participants by distributing a 1-bit or 1-qubit share to each participant. When a dealer distributes a 1-bit share to each participant, leakage of a share from 1 participant allows an adversary to gain a 1-bit classical secret and the dealer cannot distribute any share before a secret is given. On the other hand, the superdense coding [20] provides quantum secret sharing scheme, where each share is 1-qubit and a secret is 2-bits. This quantum secret sharing scheme was proposed by Gottesman [5, Section 4]. Let $I$ be the identity operator, $X$ (or $Y$ or $Z$) be a Pauli-$X$ (or $Y$ or $Z$) operator [13] and $i = \sqrt{-1}$. Gottesman's secret sharing scheme [5, Section 4] is given as follows:

1) Prepare a Bell state $|\psi\rangle = (|00\rangle + |11\rangle)/\sqrt{2}$.
2) Perform a unitary operator corresponding to a secret and prepare the quantum state of shares as follows:
   - If a secret is 00, perform $I \otimes I$ to $|\psi\rangle$ and prepare $(I \otimes I)|\psi\rangle$.
   - If a secret is 01, perform $I \otimes Z$ to $|\psi\rangle$ and prepare $(I \otimes Z)|\psi\rangle$.
   - If a secret is 10, perform $I \otimes X$ to $|\psi\rangle$ and prepare $(I \otimes X)|\psi\rangle$.
   - If a secret is 11, perform $I \otimes iY$ to $|\psi\rangle$ and prepare $(I \otimes iY)|\psi\rangle$.
3) Distribute each qubit in the quantum state of shares to each participant.

In this scheme, leakage of a share from 1 participant does not allow an adversary from gaining any information about a secret. In addition, the quantum state of shares has the form $(I \otimes U)|\psi\rangle$, where $U$ is a unitary operator for a given secret and thus a dealer can distribute the first qubit in the quantum state of shares to 1 participant before a given secret.

Matsumoto generalized Gottesman's secret sharing [5, Section 4] to an arbitrary number of participants and an arbitrary size of classical secrets [21], [22]. Matsumoto's secret sharing scheme [22] is based on quantum stabilizer codes [23], [24], [25], [26], [27], [28], [29] and summarized as follows:

1) Prepare the multipartite entangle state $|\varphi\rangle$ determined by a quantum stabilizer code.
2) Choose a unitary operator $U$ for a given secret and perform chosen unitary operator $U$ on $|\varphi\rangle$. Prepare the quantum state of shares $U|\varphi\rangle$.
3) Distribute each qudit in the quantum state of shares $U|\varphi\rangle$ to each participant.

We modify an encoding method of Matsumoto's secret sharing scheme [22] so that for any classical secrets, a dealer chooses a unitary operator of the form $U = I \otimes V$, where $I$ is the identity operator and $V$ is a unitary operator. As we can see from Gottesman's secret sharing [5, Section 4], a dealer can distribute each qudit in the quantum state of shares corresponding to an identity operator $I$ before a secret is given. Our proposed secret sharing scheme is a special case of Matsumoto's secret sharing scheme [22] and thus retains access structures of Matsumoto's secret sharing scheme [22]. In our paper, we clarify a necessary and sufficient condition on advance-shareable sets in our proposal.

Our proposal can provide a further advantage to advance sharing. In advance sharing of Fig. 1, the dealer generates a share $S_B$ based on not only a secret $M$ but also a share $S_A$. In classical secret sharing, a dealer realizes advance sharing by generating the rest of shares based on given secret and shares already distributed. Thus the dealer needs to keep shares already distributed until all shares are generated. Shares distributed in advance can leak out from the dealer. If shares distributed in advance leak out, forbidden sets are narrower than designed. It implies that smaller leakage allows an adversary to gain the partial information about a secret than designed. Thus advance sharing causes extra danger of security breaches in classical secret sharing. In our paper, we prove that extra danger of security breaches caused by advance sharing is unavoidable in any classical secret sharing. In contrast, the rest of shares are generated only based on a given secret in the proposed quantum secret sharing scheme. Thus in our proposal, the dealer does not need to keep the information about shares already distributed. For this reason, even if security breaches expose dealer's storage before a secret is given, our proposed secret sharing is as secure as designed. Therefore, our proposed secret sharing does not have extra danger of security breaches which is unavoidable in any classical secret sharing.

The paper is outlined as follows: Sect. II introduces necessary notations. In Sect. III, we modify an encoding method of Matsumoto's quantum secret sharing scheme so that some shares can be distributed before a given secret and clarify a necessary and sufficient condition on advance shareable sets in our proposal. In Sect. IV, we show a construction of our proposed encoding method from the Reed–Solomon codes and compare it with a ramp version [30] of Shamir's secret sharing. In Sect. V, we prove that extra danger of security breaches described above is unavoidable in any classical secret sharing while our proposed secret sharing has no extra danger of security breaches. In Sect. VI, we give a Gilbert–Varshamov-type sufficient condition for existence of our proposed secret sharing. We conclude our paper in Sect. VII.

## II. NOTATION

Throughout this paper, we suppose that $q = p^m$ where $p$ is a prime and $m$ is a positive integer. Denote the finite field with $q$ elements by $\mathbb{F}_q$ and the $q$-dimensional complex linear space by $\mathbb{C}_q$. The quantum state space of $n$-qudits is denoted by $\mathbb{C}_q^{\otimes n}$ with its orthonormal basis $\{|\mathbf{v}\rangle = |v_1\rangle \otimes \cdots \otimes |v_n\rangle : \mathbf{v} = (v_1, \ldots, v_n) \in \mathbb{F}_q^n\}$.

For two vectors $\mathbf{a}, \mathbf{b} \in \mathbb{F}_q^{2n}$, $\langle \mathbf{a}, \mathbf{b}\rangle_E$ denotes the standard Euclidean inner product. For two vectors $(\mathbf{a}|\mathbf{b}), (\mathbf{c}|\mathbf{d}) \in \mathbb{F}_q^{2n}$, define the standard symplectic inner product as

$$\langle (\mathbf{a}|\mathbf{b}), (\mathbf{c}|\mathbf{d})\rangle_s = \langle \mathbf{a}, \mathbf{d}\rangle_E - \langle \mathbf{c}, \mathbf{b}\rangle_E.$$

For an $\mathbb{F}_q$-linear space $C \subset \mathbb{F}_q^{2n}$, $C^{\perp s}$ denotes its orthogonal space in $\mathbb{F}_q^{2n}$ with respect to $\langle\cdot,\cdot\rangle_s$, that is,

$$C^{\perp s} = \{(\mathbf{a}|\mathbf{b}) \in \mathbb{F}_q^{2n} : \langle (\mathbf{a}|\mathbf{b}), (\mathbf{c}|\mathbf{d})\rangle_s = 0 \quad \text{for all } (\mathbf{c}|\mathbf{d}) \in C\}.$$

Let $\{\gamma_1, \ldots, \gamma_m\}$ be a fixed $\mathbb{F}_p$-basis of $\mathbb{F}_q$. Let $M$ be a $m \times m$ invertible matrix over $\mathbb{F}_p$ whose $(i,j)$ element is $\mathrm{Tr}_{q/p}(\gamma_i\gamma_j)$, where $\mathrm{Tr}_{q/p}$ is the trace map from $\mathbb{F}_q$ to $\mathbb{F}_p$. Let $\phi : \mathbb{F}_p^{2mn} \longrightarrow \mathbb{F}_q^{2n}$ be an $\mathbb{F}_p$-linear isomorphism sending $(a_{1,1}, \ldots, a_{1,m}, a_{2,1}, \ldots, a_{n,m}|b_{1,1}, \ldots, b_{1,m}, b_{2,1}, \ldots, b_{n,m})$ to

$$\left(\sum_{j=1}^m a_{1,j}\gamma_j, \ldots, \sum_{j=1}^m a_{n,j}\gamma_j \,\middle|\, \sum_{j=1}^m b'_{1,j}\gamma_j, \ldots, \sum_{j=1}^m b'_{n,j}\gamma_j\right),$$

where $(b'_{i,1}, \ldots, b'_{i,m}) = (b_{i,1}, \ldots, b_{i,m})M^{-1}$ for $i = 1, \ldots, n$. For $\alpha, \beta \in \mathbb{F}_p$, define the unitary operators $X(\alpha), Z(\beta)$ on $\mathbb{C}_p$ as

$$X(\alpha) = \sum_{x\in\mathbb{F}_p} |x+\alpha\rangle\langle x|, \quad Z(\beta) = \sum_{z\in\mathbb{F}_p} \omega^{\beta z}|z\rangle\langle z|,$$

where $\omega = e^{\frac{2\pi i}{p}}$ is a $p$-th primitive root of unity. For $(\mathbf{a}|\mathbf{b}) = (a_1, \ldots, a_n|b_1, \ldots, b_n) \in \mathbb{F}_q^{2n}$, let $\phi^{-1}(a_1, \ldots, a_n|b_1, \ldots, b_n) = (c_{1,1}, \ldots, c_{1,m}, c_{2,1}, \ldots, c_{n,m}|d_{1,1}, \ldots, d_{1,m}, d_{2,1}, \ldots, d_{n,m}) \in \mathbb{F}_p^{2mn}$ and define the unitary operator on $\mathbb{C}_q^{\otimes n}$ as

$$X(\mathbf{a})Z(\mathbf{b}) = X(c_{1,1})Z(d_{1,1}) \otimes \ldots \otimes X(c_{1,m})Z(d_{1,m}) \otimes X(c_{2,1})Z(d_{2,1}) \otimes \ldots \otimes X(c_{n,m})Z(d_{n,m}).$$

Let $C \subset \mathbb{F}_q^{2n}$ be an $\mathbb{F}_q$-linear space such that $\dim C = n-k$ and $C \subset C^{\perp s}$. An $[[n,k]]_q$ quantum stabilizer codes [23] encoding $k$ qudits into $n$ qudits can be defined as a simultaneous eigenspace of all $X(\mathbf{a})Z(\mathbf{b})((\mathbf{a}|\mathbf{b}) \in C)$.

For $(\mathbf{a}|\mathbf{b}) = (a_1, \ldots, a_n|b_1, \ldots, b_n) \in \mathbb{F}_q^{2n}$, define its symplectic weight as $\mathrm{swt}(\mathbf{a}|\mathbf{b}) = |\{i : (a_i, b_i) \neq (0,0)\}|$. For $V_2 \subset V_1 \subset \mathbb{F}_q^{2n}$, define their coset distance [31] as $d_s(V_1, V_2) = \min\{\mathrm{swt}(\mathbf{a}|\mathbf{b}) : (\mathbf{a}|\mathbf{b}) \in V_1 \setminus V_2\}$.

Let $A \subset \{1, \ldots, n\}$. Define $\mathbb{F}_q^A = \{(a_1, \ldots, a_n|b_1, \ldots, b_n) \in \mathbb{F}_q^{2n} : (a_i, b_i) = 0 \text{ for } i \notin A\}$. Let $P_A$ to be the projection map onto $A$, that is, $P_A(a_1, \ldots, a_n|b_1, \ldots, b_n) = (a_i|b_i)_{i\in A}$ for any $(a_1, \ldots, a_n|b_1, \ldots, b_n) \in \mathbb{F}_q^{2n}$.

## III. DISTRIBUTION OF QUANTUM SHARES BEFORE A GIVEN SECRET

In this section, we modify an encoding method of Matsumoto's secret sharing scheme [22] so that some shares can be distributed before a given secret.

### A. OUR PROPOSED ENCODING METHOD

Let $C_S, C_R \subset \mathbb{F}_q^{2n}$ be $\mathbb{F}_q$-linear spaces with $\dim C_S = n-k-s$, $\dim C_R = n-s$ and $C_S \subset C_R \subset C_R^{\perp s} \subset C_S^{\perp s}$. Then $C_S$ defines an $[[n, k+s]]_q$ quantum stabilizer code $\mathcal{Q}$ and $C_R$ defines an $[[n,s]]_q$ quantum stabilizer code. By Witt's lemma in [32], there always exists $C_S \subset C_R \subset C_{\max} \subset C_R^{\perp s} \subset C_S^{\perp s}$ such that $C_{\max} = C_{\max}^{\perp s}$ and $\dim C_{\max} = n$. Note that there are many possible choices of $C_{\max}$. We fix $C_{\max}$. Since $C_{\max} = C_{\max}^{\perp s}$, $C_{\max}$ defines an $[[n,0]]_q$ quantum stabilizer code $\mathcal{Q}_0$. Without loss of generality, we can assume $\mathcal{Q}_0 \subset \mathcal{Q}$. Denote a quantum state vector in $\mathcal{Q}_0$ by $|\varphi\rangle$. Since $\dim C_S^{\perp s}/C_R^{\perp s} = k$, we have an isomorphism $f : \mathbb{F}_q^k \longrightarrow C_S^{\perp s}/C_R^{\perp s}$. Note that there are many possible choices of isomorphisms $f : \mathbb{F}_q^k \longrightarrow C_S^{\perp s}/C_R^{\perp s}$. We fix $f$. Denote $n$ participants or a set of $n$ shares by $\{1, \ldots, n\}$. Let $B$ be a set of $t$ shares distributed before a given secret and $\overline{B} = \{1, \ldots, n\} \setminus B$ be the rest of shares. Without loss of generality, we can assume $B = \{1, \ldots, t\}$ and $\overline{B} = \{t+1, \ldots, n\}$.

The conventional quantum secret sharing scheme [22] using stabilizers $C_S \subset C_R \subset C_{\max}$ distributes a given classical secret $\mathbf{m} \in \mathbb{F}_q^k$ to $n$ participants as follows:

1) $f(\mathbf{m})$ is a coset of $C_S^{\perp s}/C_R^{\perp s}$ and $f(\mathbf{m})$ can also seen as a subset of $C_S^{\perp s}/C_{\max}$. Choose $T \in f(\mathbf{m}) \subset C_S^{\perp s}/C_{\max}$ at uniformly random. Choose an arbitrary $(\mathbf{a}|\mathbf{b}) \in T$ and prepare the quantum codeword $X(\mathbf{a})Z(\mathbf{b})|\varphi\rangle \in \mathcal{Q}$.
2) Distribute each qudit in the quantum codeword $X(\mathbf{a})Z(\mathbf{b})|\varphi\rangle$ to each participant.

Our proposed modified version of the conventional quantum version [22] distributes a given classical secret $\mathbf{m} \in \mathbb{F}_q^k$ to $n$ participants with a set of shares $B = \{1, \ldots, t\}$ being advance-shareable as follows:

0) Prepare the quantum codeword $|\varphi\rangle$ and distribute the $i$-th qudit in the quantum codeword $|\varphi\rangle$ to the $i$-th participant for $i \in B = \{1, \ldots, t\}$.

1) $f(\mathbf{m})$ is a coset of $C_S^{\perp s}/C_R^{\perp s}$ and $f(\mathbf{m})$ can also seen as a subset of $C_S^{\perp s}/C_{\max}$. Choose $T \in f(\mathbf{m}) \subset C_S^{\perp s}/C_{\max}$ at uniformly random. Choose a coset representative of $T$ of the form $(\mathbf{0}, \mathbf{x}|\mathbf{0}, \mathbf{y})$ with $\mathbf{x}, \mathbf{y} \in \mathbb{F}_q^{n-t}$. Prepare the quantum codeword

$$\begin{aligned} &X(\mathbf{0}, \mathbf{x})Z(\mathbf{0}, \mathbf{y})|\varphi\rangle \\ &\quad = (X(\mathbf{0})Z(\mathbf{0}) \otimes X(\mathbf{x})Z(\mathbf{y}))|\varphi\rangle \\ &\quad = (I_p^t \otimes X(\mathbf{x})Z(\mathbf{y}))|\varphi\rangle \in \mathcal{Q}, \end{aligned}$$

where $I_p^t$ is the identity operator on $\mathbb{C}_q^{\otimes t}$.

2) Distribute the $i$-th qudit in the quantum codeword $(I_p^t \otimes X(\mathbf{x})Z(\mathbf{y}))|\varphi\rangle$ to the $i$-th participant for $i \in \overline{B} = \{t+1, \ldots, n\}$.

Our proposal restricts choices of unitary operator applied by the dealer. Thus the quantum codeword has the form $X(\mathbf{0}, \mathbf{x})Z(\mathbf{0}, \mathbf{y})|\varphi\rangle = (I_p^t \otimes X(\mathbf{x})Z(\mathbf{y}))|\varphi\rangle$ for a given secret. Therefore, the dealer can distribute the $i$-th qudit in the quantum codeword before a given secret for $i \in B = \{1, \ldots, t\}$.

Focus on Step 1 of the conventional quantum version [22] and that of our proposed modified version. Assume that a dealer chooses $T \in f(\mathbf{m}) \subset C_S^{\perp s}/C_{\max}$. A dealer chooses an arbitrary coset representative $(\mathbf{a}|\mathbf{b}) \in T$ in the conventional quantum version [22] while a dealer chooses a coset representative of $T$ of the form $(\mathbf{0}, \mathbf{x}|\mathbf{0}, \mathbf{y})$ in our proposed modified version.

*Theorem 1:* Since our proposed modified version is a special case of the conventional quantum version [22], it retains access structures, and necessary and sufficient conditions on qualified sets and forbidden sets, from the conventional quantum version [22].

In our proposed secret sharing scheme, after a dealer chooses $T \in f(\mathbf{m}) \subset C_S^{\perp s}/C_{\max}$, the dealer chooses a coset representative of $T$ of the form $(\mathbf{0}, \mathbf{x}|\mathbf{0}, \mathbf{y})$. Let $(\mathbf{a}|\mathbf{b})$ be any element of $T$. We give an explicit method to compute $\mathbf{x}$, $\mathbf{y} \in \mathbb{F}_q^{n-t}$ such that $T = (\mathbf{0}, \mathbf{x}|\mathbf{0}, \mathbf{y}) + C_{\max}$ as follows: First compute a basis $\{(\mathbf{c}_1|\mathbf{d}_1), \ldots, (\mathbf{c}_n|\mathbf{d}_n)\}$ of an $\mathbb{F}_q$-linear space $C_{\max}$. Define a matrix $H$ for $C_{\max}$ as

$$H = \left[\begin{array}{c|c} \mathbf{d}_1 & -\mathbf{c}_1 \\ \mathbf{d}_2 & -\mathbf{c}_2 \\ \vdots & \vdots \\ \mathbf{d}_n & -\mathbf{c}_n \end{array}\right].$$

Next solve the following simultaneous linear equation in unknowns $\mathbf{x}$, $\mathbf{y}$ and constants $(\mathbf{a}|\mathbf{b}) \in T$:

$$([\mathbf{a}|\mathbf{b}] - [\mathbf{0}, \mathbf{x}|\mathbf{0}, \mathbf{y}])^t H = \mathbf{0}, \tag{1}$$

where ${}^tH$ represents the transpose of the matrix $H$. A solution $(\mathbf{x}, \mathbf{y})$ to (1) satisfies $T = (\mathbf{0}, \mathbf{x}|\mathbf{0}, \mathbf{y}) + C_{\max}$. If a solution to (1) exists for any $T \in C_S^{\perp s}/C_{\max}$, a set of shares $B$ is advance-shareable. There may exist $T \in C_S^{\perp s}/C_{\max}$ such that (1) has no solution. In this case, a set of shares $B$ is not advance-shareable.

Thus a set of shares $B$ is said to be advance-shareable if and only if there exists a solution to (1) for any $T \in C_S^{\perp s}/C_{\max}$. However, it is nontrivial to check if a set of shares $B$ is advance-shareable. Thus it is desirable to have a simpler condition to check if a set of shares $B$ is advance-shareable. For this reason, we will clarify a necessary and sufficient condition that a set of shares $B$ is advance-shareable in Sect. III-C.

*Remark 2:* Encoding can be classified into deterministic encoding and randomized encoding. Shares are deterministically chosen for a fixed secret in deterministic encoding, while shares are randomly chosen for a fixed secret in randomized encoding. In both conventional and modified versions, encoding is randomized and includes deterministic encoding as a special case. Encoding is deterministic if and only if $C_R = C_{\max} = C_R^{\perp s}$.

*Remark 3:* The reconstructing process of the secret in our proposal is the same as that of the conventional quantum version [22, Theorem 4].

### B. OUR PROPOSED ENCODING METHOD CONSTRUCTED FROM A SHORT LINEAR CODE

We illustrate how our proposed encoding method in Sect. III-A works with an example in [22, Example 3].

Let $q = 3, n = 4, k = s = 2, t = |B| = 2$. A basis of the doubly-extended $[4, 2, 3]_3$ Reed-Solomon code [33] over $\mathbb{F}_3$ consists of

$$\begin{aligned} \mathbf{v}_1 &= (1, 1, 1, 0), \\ \mathbf{v}_2 &= (2, 1, 0, 1). \end{aligned}$$

By using them, define $C_S = \{\mathbf{0}\}$, $C_R$ as the linear space spanned by $\{(\mathbf{v}_1|\mathbf{0}), (\mathbf{0}|\mathbf{v}_1)\}$, and $C_{\max}$ as the linear space spanned by $\{(\mathbf{v}_1|\mathbf{0}), (\mathbf{v}_2|\mathbf{0}), (\mathbf{0}|\mathbf{v}_1), (\mathbf{0}|\mathbf{v}_2)\}$. Let

$$\mathbf{v}_3 = (1, 1, 0, 0).$$

Then $C_R^{\perp s}$ is spanned by $C_{\max} \cup \{(\mathbf{v}_3|\mathbf{0}), (\mathbf{0}|\mathbf{v}_3)\}$. Let

$$\mathbf{v}_4 = (0, 0, 0, 1).$$

Then $C_S^{\perp s} = \mathbb{F}_3^8$ and $\{(\mathbf{v}_4|\mathbf{0})+C_R^{\perp s}, (\mathbf{0}|\mathbf{v}_4)+C_R^{\perp s}\}$ forms a basis of $C_S^{\perp s}/C_R^{\perp s}$. Let $|\varphi\rangle$ be an eigenvector of all unitary matrices corresponding to a vector in $C_{\max}$. Define $f : \mathbb{F}_3^2 \longrightarrow C_S^{\perp s}/C_R^{\perp s}$ as $f(m_1, m_2) = (0, 0, 0, m_1|0, 0, 0, m_2)+C_R^{\perp s} \subset \mathbb{F}_3^8$.

Let $T$ be any coset in $f(m_1, m_2) \subset C_S^{\perp s}/C_{\max}$. $C_R^{\perp s}$ is spanned by $C_{\max} \cup \{(\mathbf{v}_3|\mathbf{0}), (\mathbf{0}|\mathbf{v}_3)\}$. Thus for any $T \in f(m_1, m_2) = (0, 0, 0, m_1|0, 0, 0, m_2) + C_R^{\perp s} \subset C_S^{\perp s}/C_{\max}$, there exist $a_1, a_2 \in \mathbb{F}_3$ such that $T = (a_1, a_1, 0, m_1|a_2, a_2, 0, m_2) + C_{\max}$. An explicit derivation of $(x_1, x_2), (y_1, y_2) \in \mathbb{F}_3^2$ such that $T = (a_1, a_1, 0, m_1|a_2, a_2, 0, m_2) + C_{\max} = (0, 0, x_1, x_2|0, 0, y_1, y_2) + C_{\max}$ is given as follows: Define a matrix $H$ for $C_{\max}$ as

$$H = \left[\begin{array}{cccc|cccc} 0 & 0 & 0 & 0 & 2 & 2 & 2 & 0 \\ 1 & 1 & 1 & 0 & 0 & 0 & 0 & 0 \\ 0 & 0 & 0 & 0 & 1 & 2 & 0 & 2 \\ 2 & 1 & 0 & 1 & 0 & 0 & 0 & 0 \end{array}\right].$$

Solve the following simultaneous linear equation in unknowns $(x_1, x_2), (y_1, y_2) \in \mathbb{F}_3^2$ and constants $a_1, a_2 \in \mathbb{F}_3$:

$$\begin{aligned} &([a_1, a_1, 0, m_1|a_2, a_2, 0, m_2] \\ &\qquad -[0, 0, x_1, x_2|0, 0, y_1, y_2])^t H = \mathbf{0}. \end{aligned}$$

The solution to this equation is $(x_1, x_2) = (2a_1, m_1)$, $(y_1, y_2) = (2a_2, m_2)$. For any $T = (a_1, a_1, 0, m_1|a_2, a_2, 0, m_2)+C_{\max}$ with $a_1, a_2 \in \mathbb{F}_3$, find that

$$T = (0, 0, 2a_1, m_1|0, 0, 2a_2, m_2) + C_{\max}.$$

A dealer can perform the following secret sharing scheme for a given classical secret $(m_1, m_2) \in \mathbb{F}_3^2$ to 4 participants with a set of shares $\{1, 2\}$ being advance-shareable:

0) Prepare the quantum codeword $|\varphi\rangle$ and distribute the $i$-th qudit in the quantum codeword $|\varphi\rangle$ to the $i$-th participant for $i \in \{1, 2\}$.
1) $f(m_1, m_2)$ is a coset of $C_S^{\perp s}/C_R^{\perp s}$ and $f(m_1, m_2)$ can also seen as a subset of $C_S^{\perp s}/C_{\max}$. Choose $T = (a_1, a_1, 0, m_1|a_2, a_2, 0, m_2) + C_{\max} \in f(m_1, m_2) \subset C_S^{\perp s}/C_{\max}$ at uniformly random, with $a_1, a_2 \in \mathbb{F}_3$. This means that the dealer chooses $a_1, a_2 \in \mathbb{F}_3$ at uniformly random and fix $a_1, a_2 \in \mathbb{F}_3$. Choose a coset representative of the form $(0, 0, 2a_1, m_1|0, 0, 2a_2, m_2)$. Prepare the quantum codeword

$$\begin{aligned} &X(0, 0, 2a_1, m_1)Z(0, 0, 2a_2, m_2)|\varphi\rangle \\ &= (X(0, 0)Z(0, 0) \otimes X(2a_1, m_1)Z(2a_2, m_2))|\varphi\rangle \\ &= (I_p^2 \otimes X(2a_1, m_1)Z(2a_2, m_2))|\varphi\rangle \in \mathcal{Q}, \end{aligned}$$

where $I_p^2$ is the identity operator on $\mathbb{C}_p^{\otimes 2}$.
2) Distribute the $i$-th qudit in the quantum codeword $(I_p^2 \otimes X(2a_1, m_1)Z(2a_2, m_2))|\varphi\rangle$ to the $i$-th participant for $i \in \{3, 4\}$.

In this scheme, $C_S, C_R, C_{\max}$ are defined as the same as [22, Example 3]. Thus by Theorem 1, this scheme retains the access structure of the conventional quantum secret sharing scheme [22, Example 3]. That is, a set of 2 or less than 2 shares is forbidden and the set of all 4 shares is qualified.

### C. NECESSARY AND SUFFICIENT CONDITION ON ADVANCE-SHAREABLE SETS

We clarify a necessary and sufficient condition that a set of shares $B$ is advance-shareable in our proposal.

*Theorem 4:* A set of shares $B$ is advance-shareable or equivalently for any $T \in C_S^{\perp s}/C_{\max}$, there exists $(\mathbf{z}|\mathbf{w}) \in C_S^{\perp s} \cap \mathbb{F}_q^{\overline{B}}$ such that $T = (\mathbf{z}|\mathbf{w}) + C_{\max}$ if and only if

$$\dim C_S^{\perp s} \cap \mathbb{F}_q^{\overline{B}}/C_{\max} \cap \mathbb{F}_q^{\overline{B}} = \dim C_S^{\perp s}/C_{\max}.$$

To prove Theorem 4, we give several lemmas. Theorem 4 immediately follows from Lemma 5 and Lemma 7.

*Lemma 5:* The following conditions (i) and (ii) are equivalent:

(i) For any coset $T \in C_S^{\perp s}/C_{\max}$, there exists $(\mathbf{z}|\mathbf{w}) \in C_S^{\perp s} \cap \mathbb{F}_q^{\overline{B}}$ such that $T = (\mathbf{z}|\mathbf{w}) + C_{\max}$.
(ii) $C_S^{\perp s} = (C_S^{\perp s} \cap \mathbb{F}_q^{\overline{B}}) + C_{\max}$, where $(C_S^{\perp s} \cap \mathbb{F}_q^{\overline{B}}) + C_{\max}$ is an $\mathbb{F}_q$-linear space spanned by an $\mathbb{F}_q$-linear spaces $C_S^{\perp s} \cap \mathbb{F}_q^{\overline{B}}$ and $C_{\max}$.

*Proof:* Assume that the condition (i) holds. Let $(\mathbf{a}|\mathbf{b})$ be any element of $C_S^{\perp s}$. Then there exists a coset $T \in C_S^{\perp s}/C_{\max}$ such that $(\mathbf{a}|\mathbf{b}) \in T$. By the condition (i), there exists $(\mathbf{z}|\mathbf{w}) \in C_S^{\perp s} \cap \mathbb{F}_q^{\overline{B}}$ such that $T = (\mathbf{z}|\mathbf{w}) + C_{\max}$. This means that there exist $(\mathbf{z}|\mathbf{w}) \in C_S^{\perp s} \cap \mathbb{F}_q^{\overline{B}}$ and $(\mathbf{c}|\mathbf{d}) \in C_{\max}$ such that $(\mathbf{a}|\mathbf{b}) = (\mathbf{z}|\mathbf{w}) + (\mathbf{c}|\mathbf{d})$. This shows that $C_S^{\perp s} \subset (C_S^{\perp s} \cap \mathbb{F}_q^{\overline{B}}) + C_{\max}$. Obviously, $C_S^{\perp s} \supset (C_S^{\perp s} \cap \mathbb{F}_q^{\overline{B}}) + C_{\max}$ holds. Therefore, the condition (ii) holds.

Assume that the condition (ii) holds. Let $T$ be any coset of $C_S^{\perp s}/C_{\max}$. Then there exists $(\mathbf{a}|\mathbf{b}) \in C_S^{\perp s}$ such that $T = (\mathbf{a}|\mathbf{b}) + C_{\max}$. Since $(\mathbf{a}|\mathbf{b}) \in C_S^{\perp s} = (C_S^{\perp s} \cap \mathbb{F}_q^{\overline{B}}) + C_{\max}$, there exist $(\mathbf{z}|\mathbf{w}) \in C_S^{\perp s} \cap \mathbb{F}_q^{\overline{B}}$, $(\mathbf{c}|\mathbf{d}) \in C_{\max}$ such that $(\mathbf{a}|\mathbf{b}) = (\mathbf{z}|\mathbf{w}) + (\mathbf{c}|\mathbf{d})$. This shows that $T = (\mathbf{a}|\mathbf{b}) + C_{\max} = (\mathbf{z}|\mathbf{w})+(\mathbf{c}|\mathbf{d})+C_{\max} = (\mathbf{z}|\mathbf{w})+C_{\max}$. Therefore, the condition (i) holds. ■

*Lemma 6:* Let $V, W$ be any $\mathbb{F}_q$-linear spaces such that $V \subset W \subset C_{\max} = C_{\max}^{\perp s} \subset W^{\perp s} \subset V^{\perp s}$. Define an $\mathbb{F}_q$-linear map $\psi : V^{\perp s} \cap \mathbb{F}_q^{\overline{B}}/W^{\perp s} \cap \mathbb{F}_q^{\overline{B}} \longrightarrow ((V^{\perp s} \cap \mathbb{F}_q^{\overline{B}}) + C_{\max})/((W^{\perp s} \cap \mathbb{F}_q^{\overline{B}}) + C_{\max})$ as $\psi((\mathbf{a}|\mathbf{b}) + W^{\perp s} \cap \mathbb{F}_q^{\overline{B}}) = (\mathbf{a}|\mathbf{b}) + (W^{\perp s} \cap \mathbb{F}_q^{\overline{B}}) + C_{\max}$ for any $(\mathbf{a}|\mathbf{b}) \in V^{\perp s} \cap \mathbb{F}_q^{\overline{B}}$. Then $\psi : V^{\perp s} \cap \mathbb{F}_q^{\overline{B}}/W^{\perp s} \cap \mathbb{F}_q^{\overline{B}} \longrightarrow ((V^{\perp s} \cap \mathbb{F}_q^{\overline{B}}) + C_{\max})/((W^{\perp s} \cap \mathbb{F}_q^{\overline{B}}) + C_{\max})$ is an isomorphism.

*Proof:* Assume that $(\mathbf{a}|\mathbf{b})+W^{\perp s}\cap\mathbb{F}_q^{\overline{B}} = (\mathbf{c}|\mathbf{d})+W^{\perp s}\cap\mathbb{F}_q^{\overline{B}}$ for $(\mathbf{a}|\mathbf{b}), (\mathbf{c}|\mathbf{d}) \in V^{\perp s} \cap \mathbb{F}_q^{\overline{B}}$. Then we have $(\mathbf{a}|\mathbf{b}) - (\mathbf{c}|\mathbf{d}) \in W^{\perp s} \cap \mathbb{F}_q^{\overline{B}} \subset (W^{\perp s} \cap \mathbb{F}_q^{\overline{B}}) + C_{\max}$. This means that $\psi((\mathbf{a}|\mathbf{b}) + W^{\perp s} \cap \mathbb{F}_q^{\overline{B}}) = \psi((\mathbf{c}|\mathbf{d}) + W^{\perp s} \cap \mathbb{F}_q^{\overline{B}})$. This shows that $\psi$ is well-defined.

Assume that $\psi((\mathbf{a}|\mathbf{b})+W^{\perp s}\cap\mathbb{F}_q^{\overline{B}}) = (\mathbf{a}|\mathbf{b})+(W^{\perp s}\cap\mathbb{F}_q^{\overline{B}})+C_{\max} = (W^{\perp s} \cap \mathbb{F}_q^{\overline{B}}) + C_{\max}$ for $(\mathbf{a}|\mathbf{b}) \in V^{\perp s} \cap \mathbb{F}_q^{\overline{B}}$. Then we have $(\mathbf{a}|\mathbf{b}) \in (W^{\perp s} \cap \mathbb{F}_q^{\overline{B}}) + C_{\max}$. This means that there exist $(\mathbf{c}|\mathbf{d}) \in W^{\perp s} \cap \mathbb{F}_q^{\overline{B}}$ and $(\mathbf{e}|\mathbf{f}) \in C_{\max}$ such that $(\mathbf{a}|\mathbf{b}) = (\mathbf{c}|\mathbf{d})+(\mathbf{e}|\mathbf{f})$. Since $(\mathbf{a}|\mathbf{b}) \in V^{\perp s}\cap\mathbb{F}_q^{\overline{B}}$, we have $(\mathbf{e}|\mathbf{f}) = (\mathbf{a}|\mathbf{b})-(\mathbf{c}|\mathbf{d}) \in C_{\max}\cap\mathbb{F}_q^{\overline{B}}$. Therefore, we have $(\mathbf{a}|\mathbf{b}) = (\mathbf{c}|\mathbf{d})+(\mathbf{e}|\mathbf{f}) \in W^{\perp s}\cap\mathbb{F}_q^{\overline{B}}$. This shows that the kernel of $\psi$ is zero and thus an $\mathbb{F}_q$-linear map $\psi$ is injective.

Let $(\mathbf{n}|\mathbf{m})+(W^{\perp s}\cap\mathbb{F}_q^{\overline{B}})+C_{\max}$ be any element of $((V^{\perp s} \cap \mathbb{F}_q^{\overline{B}})+C_{\max})/((W^{\perp s}\cap\mathbb{F}_q^{\overline{B}})+C_{\max})$. Since $(\mathbf{n}|\mathbf{m}) \in (V^{\perp s}\cap\mathbb{F}_q^{\overline{B}})+C_{\max}$, there exist $(\mathbf{a}|\mathbf{b}) \in V^{\perp s} \cap \mathbb{F}_q^{\overline{B}}$ and $(\mathbf{c}|\mathbf{d}) \in C_{\max}$ such that $(\mathbf{n}|\mathbf{m}) = (\mathbf{a}|\mathbf{b}) + (\mathbf{c}|\mathbf{d})$. Since $(\mathbf{c}|\mathbf{d}) \in C_{\max} \subset (W^{\perp s} \cap \mathbb{F}_q^{\overline{B}}) + C_{\max}$, we have $(\mathbf{n}|\mathbf{m}) + (W^{\perp s} \cap \mathbb{F}_q^{\overline{B}}) + C_{\max} = (\mathbf{a}|\mathbf{b}) + (W^{\perp s} \cap \mathbb{F}_q^{\overline{B}}) + C_{\max}$. Namely, there exists $(\mathbf{a}|\mathbf{b}) \in V^{\perp s} \cap \mathbb{F}_q^{\overline{B}}$ such that $\psi((\mathbf{a}|\mathbf{b})+W^{\perp s}\cap\mathbb{F}_q^{\overline{B}}) = (\mathbf{n}|\mathbf{m})+(W^{\perp s}\cap\mathbb{F}_q^{\overline{B}})+C_{\max}$. This shows that $\psi$ is surjective.

Therefore, $\psi : V^{\perp s} \cap \mathbb{F}_q^{\overline{B}}/W^{\perp s} \cap \mathbb{F}_q^{\overline{B}} \longrightarrow ((V^{\perp s} \cap \mathbb{F}_q^{\overline{B}}) + C_{\max})/((W^{\perp s} \cap \mathbb{F}_q^{\overline{B}}) + C_{\max})$ is an isomorphism. ■

*Lemma 7:*

$$\dim C_S^{\perp s} \cap \mathbb{F}_q^{\overline{B}}/C_{\max} \cap \mathbb{F}_q^{\overline{B}} = \dim C_S^{\perp s}/C_{\max} \tag{2}$$

if and only if

$$C_S^{\perp s} = (C_S^{\perp s} \cap \mathbb{F}_q^{\overline{B}}) + C_{\max} \tag{3}$$

*Proof:* Assume that (2) holds. We have

$$\begin{aligned}&\dim C_{\mathrm{S}}^{\perp s}/C_{\max} && (4)\\ &= \dim C_{\mathrm{S}}^{\perp s} \cap \mathbb{F}_q^{\overline{B}}/C_{\max} \cap \mathbb{F}_q^{B} && (5)\\ &= \dim((C_{\mathrm{S}}^{\perp s} \cap \mathbb{F}_q^{\overline{B}}) + C_{\max})/C_{\max}, && (6)\end{aligned}$$

where (5) follows from (2), and (6) follows from $(C_{\max}^{\perp s} \cap \mathbb{F}_q^{\overline{B}}) + C_{\max} = C_{\max}$ and Lemma 6. This means that $\dim C_{\mathrm{S}}^{\perp s} = \dim(C_{\mathrm{S}}^{\perp s} \cap \mathbb{F}_q^{\overline{B}}) + C_{\max}$. Since obviously $C_{\mathrm{S}}^{\perp s} \supset (C_{\mathrm{S}}^{\perp s} \cap \mathbb{F}_q^{\overline{B}}) + C_{\max}$ holds, we have $C_{\mathrm{S}}^{\perp s} = (C_{\mathrm{S}}^{\perp s} \cap \mathbb{F}_q^{\overline{B}}) + C_{\max}$. Thus (3) holds.

Assume that (3) holds. We have

$$\begin{aligned}&\dim C_{\mathrm{S}}^{\perp s} \cap \mathbb{F}_q^{\overline{B}}/C_{\max} \cap \mathbb{F}_q^{B}\\ &= \dim((C_{\mathrm{S}}^{\perp s} \cap \mathbb{F}_q^{\overline{B}}) + C_{\max})/C_{\max} && (7)\\ &= \dim C_{\mathrm{S}}^{\perp s}/C_{\max}, && (8)\end{aligned}$$

where (7) follows from $(C_{\max}^{\perp s} \cap \mathbb{F}_q^{\overline{B}}) + C_{\max} = C_{\max}$ and Lemma 6, and (8) follows from (3). Thus (2) holds. ∎

## D. SUFFICIENT CONDITION ON ADVANCE-SHAREABLE SETS

We present a sufficient condition that a set of shares $B$ is advance-shareable in our proposal. The following sufficient condition can be verified without computing dimensions of linear spaces like Theorem 4.

*Theorem 8:* If $|B| \le d_s(C_{\max}, C_{\mathrm{S}}) - 1$, then a set of shares $B$ is advance-shareable or equivalently $\dim C_{\mathrm{S}}^{\perp s} \cap \mathbb{F}_q^{\overline{B}}/C_{\max} \cap \mathbb{F}_q^{B} = \dim C_{\mathrm{S}}^{\perp s}/C_{\max}$.

*Proof:* By the condition $|B| \le d_s(C_{\max}, C_{\mathrm{S}}) - 1$, there is no $(\mathbf{a}|\mathbf{b}) \in C_{\max} \cap \mathbb{F}_q^{B} \setminus C_{\mathrm{S}} \cap \mathbb{F}_q^{B}$. This means that

$$\dim C_{\mathrm{S}} \cap \mathbb{F}_q^{B} = \dim C_{\max} \cap \mathbb{F}_q^{B}. \quad (9)$$

By [33], [34], we have

$$\dim C_{\mathrm{S}}^{\perp s} \cap \mathbb{F}_q^{\overline{B}}/C_{\max} \cap \mathbb{F}_q^{B} = \dim P_{\overline{B}}(C_{\max})/P_{\overline{B}}(C_{\mathrm{S}}). \quad (10)$$

Since the kernel of $P_{\overline{B}} : C_{\mathrm{S}} \longrightarrow P_{\overline{B}}(C_{\mathrm{S}})$ is $C_{\mathrm{S}} \cap \mathbb{F}_q^{B}$, we have

$$\dim C_{\mathrm{S}} - \dim C_{\mathrm{S}} \cap \mathbb{F}_q^{B} = \dim P_{\overline{B}}(C_{\mathrm{S}}). \quad (11)$$

Since the kernel of $P_{\overline{B}} : C_{\max} \longrightarrow P_{\overline{B}}(C_{\max})$ is $C_{\max} \cap \mathbb{F}_q^{B}$, we have

$$\dim C_{\max} - \dim C_{\max} \cap \mathbb{F}_q^{B} = \dim P_{\overline{B}}(C_{\max}). \quad (12)$$

Therefore, we find that

$$\begin{aligned}&\dim C_{\mathrm{S}}^{\perp s} \cap \mathbb{F}_q^{\overline{B}}/C_{\max} \cap \mathbb{F}_q^{B}\\ &= \dim P_{\overline{B}}(C_{\max})/P_{\overline{B}}(C_{\mathrm{S}}) && (13)\\ &= \dim P_{\overline{B}}(C_{\max}) - \dim P_{\overline{B}}(C_{\mathrm{S}})\\ &= (\dim C_{\max} - \dim C_{\max} \cap \mathbb{F}_q^{B})\\ &\quad -(\dim C_{\mathrm{S}} - \dim C_{\mathrm{S}} \cap \mathbb{F}_q^{B}) && (14)\\ &= \dim C_{\max} - \dim C_{\mathrm{S}} && (15)\\ &= \dim C_{\mathrm{S}}^{\perp s}/C_{\max},\end{aligned}$$

where (13) follows from (10), (14) follows from (11) and (12), and (15) follows from (9). ∎

## E. RELATIONSHIP BETWEEN ADVANCE-SHAREABLE SETS AND FORBIDDEN SETS

We clarify a relationship between advance-shareable sets and forbidden sets in our proposal.

By Theorem 1, a necessary and sufficient condition on forbidden sets remains the same as the conventional quantum secret sharing scheme [22]. Thus by [22, Theorem 4], a set of shares $B$ is forbidden if and only if

$$\dim C_{\mathrm{R}} \cap \mathbb{F}_q^{B}/C_{\mathrm{S}} \cap \mathbb{F}_q^{B} = 0.$$

*Theorem 9:* If a set of shares $B$ is advance-shareable, then a set of shares $B$ is forbidden. Equivalently, if

$$\dim C_{\mathrm{S}}^{\perp s} \cap \mathbb{F}_q^{\overline{B}}/C_{\max} \cap \mathbb{F}_q^{B} = \dim C_{\mathrm{S}}^{\perp s}/C_{\max}, \quad (16)$$

then

$$\dim C_{\mathrm{R}} \cap \mathbb{F}_q^{B}/C_{\mathrm{S}} \cap \mathbb{F}_q^{B} = 0. \quad (17)$$

*Proof:* Assume (16) holds. Let $T$ be any coset in $C_{\mathrm{R}}^{\perp s}/C_{\max}$. Then we have $T \in C_{\mathrm{R}}^{\perp s}/C_{\max} \subset C_{\mathrm{S}}^{\perp s}/C_{\max}$. By (16) and Theorem 4, there exists $(\mathbf{z}|\mathbf{w}) \in C_{\mathrm{S}}^{\perp s} \cap \mathbb{F}_q^{\overline{B}}$ such that $T = (\mathbf{z}|\mathbf{w}) + C_{\max}$. Since $T \in C_{\mathrm{R}}^{\perp s}/C_{\max}$, we have $(\mathbf{z}|\mathbf{w}) \in (C_{\mathrm{S}}^{\perp s} \cap \mathbb{F}_q^{\overline{B}}) \cap C_{\mathrm{R}}^{\perp s} = C_{\mathrm{R}}^{\perp s} \cap \mathbb{F}_q^{\overline{B}}$. Thus for any $T \in C_{\mathrm{R}}^{\perp s}/C_{\max}$, there exists $(\mathbf{z}|\mathbf{w}) \in C_{\mathrm{R}}^{\perp s} \cap \mathbb{F}_q^{\overline{B}}$ such that $T = (\mathbf{z}|\mathbf{w}) + C_{\max}$. By replacing $C_{\mathrm{S}}$ with $C_{\mathrm{R}}$ in Lemma 5, we have $C_{\mathrm{R}}^{\perp s} = (C_{\mathrm{R}}^{\perp s} \cap \mathbb{F}_q^{\overline{B}}) + C_{\max}$. By (16) and Lemma 7, we have $C_{\mathrm{S}}^{\perp s} = (C_{\mathrm{S}}^{\perp s} \cap \mathbb{F}_q^{\overline{B}}) + C_{\max}$. Therefore, we have

$$\begin{aligned}&\dim C_{\mathrm{S}}^{\perp s} \cap \mathbb{F}_q^{\overline{B}}/C_{\mathrm{R}}^{\perp s} \cap \mathbb{F}_q^{\overline{B}}\\ &= \dim((C_{\mathrm{S}}^{\perp s} \cap \mathbb{F}_q^{\overline{B}}) + C_{\max})/((C_{\mathrm{R}}^{\perp s} \cap \mathbb{F}_q^{\overline{B}}) + C_{\max}) && (18)\\ &= \dim C_{\mathrm{S}}^{\perp s}/C_{\mathrm{R}}^{\perp s} && (19)\\ &= \dim C_{\mathrm{R}}/C_{\mathrm{S}}, && (20)\end{aligned}$$

where (18) follows from Lemma 6, and (19) follows from $C_{\mathrm{S}}^{\perp s} = (C_{\mathrm{S}}^{\perp s} \cap \mathbb{F}_q^{\overline{B}}) + C_{\max}$ and $C_{\mathrm{R}}^{\perp s} = (C_{\mathrm{R}}^{\perp s} \cap \mathbb{F}_q^{\overline{B}}) + C_{\max}$. Let $P_{\overline{B}}$ to be the projection map onto $\overline{B}$. Since the kernel of $P_{\overline{B}} : C_{\mathrm{S}} \longrightarrow P_{\overline{B}}(C_{\mathrm{S}})$ is $C_{\mathrm{S}} \cap \mathbb{F}_q^{B}$, we have

$$\dim C_{\mathrm{S}} - \dim C_{\mathrm{S}} \cap \mathbb{F}_q^{B} = \dim P_{\overline{B}}(C_{\mathrm{S}}). \quad (21)$$

Since the kernel of $P_{\overline{B}} : C_{\mathrm{R}} \longrightarrow P_{\overline{B}}(C_{\mathrm{R}})$ is $C_{\mathrm{R}} \cap \mathbb{F}_q^{B}$, we have

$$\dim C_{\mathrm{R}} - \dim C_{\mathrm{R}} \cap \mathbb{F}_q^{B} = \dim P_{\overline{B}}(C_{\mathrm{R}}). \quad (22)$$

By [33], [34], we have

$$\dim C_{\mathrm{S}}^{\perp s} \cap \mathbb{F}_q^{\overline{B}}/C_{\mathrm{R}}^{\perp s} \cap \mathbb{F}_q^{\overline{B}} = \dim P_{\overline{B}}(C_{\mathrm{R}})/P_{\overline{B}}(C_{\mathrm{S}}). \quad (23)$$

Therefore, we obtain (17) as follows:

$$\begin{aligned}&\dim C_{\mathrm{R}} \cap \mathbb{F}_q^{B}/C_{\mathrm{S}} \cap \mathbb{F}_q^{B}\\ &= \dim C_{\mathrm{R}} \cap \mathbb{F}_q^{B} - \dim C_{\mathrm{S}} \cap \mathbb{F}_q^{B}\\ &= (\dim C_{\mathrm{R}} - \dim P_{\overline{B}}(C_{\mathrm{R}}))\\ &\quad -(\dim C_{\mathrm{S}} - \dim P_{\overline{B}}(C_{\mathrm{S}})) && (24)\\ &= \dim C_{\mathrm{R}} - \dim C_{\mathrm{S}} - \dim P_{\overline{B}}(C_{\mathrm{R}}))/P_{\overline{B}}(C_{\mathrm{S}})) \\ &= \dim C_{\mathrm{R}} - \dim C_{\mathrm{S}} - \dim C_{\mathrm{S}}^{\perp s} \cap \mathbb{F}_q^{\overline{B}}/C_{\mathrm{R}}^{\perp s} \cap \mathbb{F}_q^{\overline{B}} && (25)\end{aligned}$$

$$= \dim C_{\mathrm{R}} - \dim C_{\mathrm{S}} - \dim C_{\mathrm{R}}/C_{\mathrm{S}} \tag{26}$$
$$= 0,$$

where (24) follows from (21) and (22), and (25) follows from (23), and (26) follows from (20). ■

## IV. OUR PROPOSED ENCODING METHOD CONSTRUCTED FROM THE REED–SOLOMON CODES

In this section, we show a construction of our proposed secret sharing scheme by using the Reed–Solomon codes in which we can easily see from $|A|$ if a set of shares $A$ is qualified, forbidden and advance-shareable. In this scheme, $C_{\mathrm{S}}$, $C_{\mathrm{R}}$, $C_{\max}$ are defined as the same as [22, Sect. 6.2]. Thus by Theorem 1, this scheme retains the access structure of the conventional quantum secret sharing scheme [22, Sect. 6.2].

Let $n = q$ and let $n - s$, $k$ be positive even integers which implies that $n - k - s$, $n + s$, $n + k + s$ are even. Let $\alpha_1$, $\ldots$, $\alpha_n \in \mathbb{F}_q$ be $n$ distinct elements. Define an $[n, k]$ Reed-Solomon(RS) code [33] as

$$\mathrm{RS}(n, k) = \{(g(\alpha_1), \ldots, g(\alpha_n)) : g(x) \in \mathbb{F}_q[x], \deg g(x) < k\}.$$

Denote the Euclidean dual by "$\perp E$". Then $\mathrm{RS}(n, k)^{\perp E} = \mathrm{RS}(n, n-k)$ [33] because $n = q$.

Define

$$C_{\mathrm{S}} = \{(\mathbf{a}|\mathbf{b}) : \mathbf{a}, \mathbf{b} \in \mathrm{RS}(n, (n-k-s)/2)\},$$
$$C_{\mathrm{R}} = \{(\mathbf{a}|\mathbf{b}) : \mathbf{a}, \mathbf{b} \in \mathrm{RS}(n, (n-s)/2)\}.$$

Then we find that

$$C_{\mathrm{R}}^{\perp s} = \{(\mathbf{a}|\mathbf{b}) : \mathbf{a}, \mathbf{b} \in \mathrm{RS}(n, (n+s)/2)\},$$
$$C_{\mathrm{S}}^{\perp s} = \{(\mathbf{a}|\mathbf{b}) : \mathbf{a}, \mathbf{b} \in \mathrm{RS}(n, (n+k+s)/2)\},$$
$$\dim C_{\mathrm{S}} = n - k - s,$$
$$\dim C_{\mathrm{R}} = n - s.$$

We can choose $C_{\max}$ as

$$C_{\max} = \{(\mathbf{a}|\mathbf{b}) : \mathbf{a} \in \mathrm{RS}(n, \lfloor n/2 \rfloor), \mathbf{b} \in \mathrm{RS}(n, \lceil n/2 \rceil)\}.$$

Let $(\mathbf{a}|\mathbf{b})$ be any element of $C_{\max}$. Then we have $\mathbf{a} \in \mathrm{RS}(n, \lfloor n/2 \rfloor)$ and $\mathbf{b} \in \mathrm{RS}(n, \lceil n/2 \rceil)$. There exists a polynomial $g(x) \in \mathbb{F}_q[x]$ such that $\deg g(x) < \lfloor n/2 \rfloor$ and $\mathbf{a} = (g(\alpha_1), \ldots, g(\alpha_n))$. There exists a polynomial $h(x) \in \mathbb{F}_q[x]$ such that $\deg h(x) < \lceil n/2 \rceil$ and $\mathbf{b} = (h(\alpha_1), \ldots, h(\alpha_n))$. If $g(x), h(x) \neq 0$, the number of distinct roots of $g(x)$ is less than $\lfloor n/2 \rfloor$ and the number of distinct roots of $h(x)$ less than $\lceil n/2 \rceil$. Then we have $\mathrm{swt}(\mathbf{a}|\mathbf{b}) = |\{i : (a_i, b_i) \neq (0, 0)\}| \geq n - \lfloor n/2 \rfloor + 1 \geq \lceil n/2 \rceil + 1$. Thus we have $d_s(C_{\max}, C_{\mathrm{S}}) - 1 \geq d_s(C_{\max}, \{\mathbf{0}\}) - 1 \geq \lceil n/2 \rceil$. Thus a set of $\lceil n/2 \rceil$ shares is advance-shareable by Theorem 8.

Therefore, we can easily see from $|A|$ if a set of shares $A$ is qualified, forbidden and advance-shareable as follows:

- A set of shares $A$ is forbidden if and only if
$$0 \leq |A| \leq \frac{n+s}{2}. \tag{27}$$
- A set of shares $A$ is qualified if and only if
$$\frac{n+k+s}{2} \leq |A| \leq n. \tag{28}$$
- A set of shares $A$ is advance-shareable if
$$0 \leq |A| \leq \left\lceil \frac{n}{2} \right\rceil. \tag{29}$$

We will compare the size of forbidden sets and advance-shareable sets of our proposed quantum secret sharing scheme constructed from the Reed–Solomon codes with that of a classical linear secret sharing. To the best of the authors' knowledge, a necessary and sufficient condition on advance-shareable sets in classical linear secret sharing has not been studied by other researchers to date. For this reason, we clarify a necessary and sufficient condition on advance-shareable sets in classical linear secret sharing.

First we give a definition of encoding of classical linear secret sharing schemes [35], [36], [37].

*Definition 10:* Let $\{1, \ldots, n\}$ be $n$ participants or equivalently a set of $n$ shares. Let $C_1 \subset \mathbb{F}_q^n$ be a linear code and $C_2 \subset C_1$ be a subcode of $C_1$. Denote $k = \dim C_1/C_2$. Fix an arbitrary isomorphism $g : \mathbb{F}_q^k \longrightarrow C_1/C_2$. Let $\mathbf{m} \in \mathbb{F}_q^k$ be a secret. From a coset $g(\mathbf{m}) \in C_1/C_2$, a dealer chooses a vector $\mathbf{c} = (c_1, \ldots, c_n) \in g(\mathbf{m}) \subset C_1$ at uniformly random. Then the dealer distributes each element $c_i$ to the $i$-th participant. It was shown in [37] that encoding in any classical linear secret sharing can be written as Definition 10.

Next we clarify a necessary and sufficient condition on advance-shareable sets of shares in classical linear secret sharing.

*Lemma 11:* In classical linear secret sharing, a set of shares $A$ is advance-shareable if and only if $A$ is forbidden.

*Proof:* Assume that a set of shares $A$ is forbidden, which is necessary and sufficient to $P_A(C_1) = P_A(C_2)$ [35], [36], [37], where $P_A$ is the projection map onto $A$. Let $\mathbf{y}$ be any element of $P_A(C_1)$. For any $g(\mathbf{m}) \in C_1/C_2$, there exists $\mathbf{x} \in g(\mathbf{m})$ such that $P_A(\mathbf{x}) = \mathbf{y}$. Thus a dealer can distribute a set of shares $A$ before a given secret.

Assume that a set of shares $A$ is not forbidden, which is necessary and sufficient to $P_A(C_1) \neq P_A(C_2)$ [35], [36], [37]. Let $\mathbf{y}$ be any element of $P_A(C_1) \setminus P_A(C_2)$. For a given secret $\mathbf{m}$ such that $g(\mathbf{m}) = \mathbf{0} + C_2 \in C_1/C_2$, there does not exist $\mathbf{x} \in g(\mathbf{m})$ such that $P_A(\mathbf{x}) = \mathbf{y}$. Thus the dealer cannot distribute a set of shares $A$ before a given secret. ■

Finally we compare the size of forbidden sets and advance-shareable sets of our proposed quantum secret sharing scheme constructed from the Reed–Solomon codes with that of a classical linear secret sharing. In our proposed quantum secret sharing scheme, a dealer shares a $k \log_2 q$-bit secret (i.e., a secret is an element of $\mathbb{F}_q^k$) to $n$ participants by distributing a $\log_2 q$-qubit share (i.e., $q$-qudit share) to each participant. Consider a ramp version [30] of Shamir's scheme as a classical linear secret sharing. Assume that in a ramp version [30] of Shamir's scheme,

- A classical secret is shared between $n$ participants.

**TABLE 1.** Comparison between our proposed quantum secret sharing scheme and ramp Shamir's scheme [30].

| | Proposed quantum secret sharing scheme | Ramp Shamir's scheme [30] |
|---|---|---|
| The bit-size of a secret | $k\log_2 q$-bit | $k\log_2 q$-bit |
| The bit-size of each share | $\log_2 q$-bit | $\log_2 q$-qubit |
| The size of qualified sets $A$ | $\frac{n+k+s}{2} \leq \lvert A\rvert \leq n$ | $\frac{n+k+s}{2} \leq \lvert A\rvert \leq n$ |
| The size of forbidden sets $A$ | $0 \leq \lvert A\rvert \leq \frac{n+s}{2}$ | $0 \leq \lvert A\rvert \leq \frac{n+s-k}{2}$ |
| The size of advance-shareable sets $A$ | $0 \leq \lvert A\rvert \leq \lceil \frac{n}{2} \rceil$ | $0 \leq \lvert A\rvert \leq \frac{n+s-k}{2}$ |

- The bit-size of a shared classical secret is $k\log_2 q$-bit (i.e., a secret is an element of $\mathbb{F}_q^k$).
- The bit-size of each share is $\log_2 q$-bit (i.e., a share is an element of $\mathbb{F}_q$).
- A set of shares $A$ is qualified if and only if

$$\frac{n+k+s}{2} \leq |A| \leq n, \tag{30}$$

which implies that the size of qualified set is the same as (28), that is, that of our proposed quantum secret sharing constructed from the Reed–Solomon code.

The bit-size of a secret and each share and the size of qualified sets, forbidden sets and advance-shareable sets in our proposed quantum secret sharing scheme constructed from the Reed–Solomon codes and a ramp version [30] of Shamir's scheme can be summarized in Table 1.

Since in a ramp version [30] of Shamir's scheme, the size of qualified sets is given by (28) and the size of each share is $1/k$ of the bit-size of the secret, a set of shares $A$ is forbidden if and only if

$$0 \leq |A| \leq \frac{n+k+s}{2} - k = \frac{n+s-k}{2}. \tag{31}$$

By Lemma 11, a set of shares $A$ is advance-shareable if and only if a set of shares $A$ is forbidden. Thus a set of shares $A$ is advance-shareable if and only if

$$0 \leq |A| \leq \frac{n+s-k}{2}. \tag{32}$$

Forbidden sets of a ramp version [30] of Shamir's scheme are narrower than that of our proposed quantum secret sharing scheme constructed from the Reed–Solomon codes, which follows from (27) and (31). It implies that smaller leakage allows an adversary to gain the partial information about a secret than our proposed quantum secret sharing scheme constructed from the Reed–Solomon codes. Our proposed quantum secret sharing scheme constructed from the Reed–Solomon codes can make the size of advance-shareable sets larger than a ramp version [30] of Shamir's scheme when $s < k$. This follows from (29) and (32).

## V. UNAVOIDABLE EXTRA DANGER OF SECURITY BREACHES IN ANY CLASSICAL SECRET SHARING WITH ADVANCE SHARING

In this section, we prove that extra danger of security breaches caused by advance sharing is unavoidable in any classical secret sharing while our proposed secret sharing does not have any extra danger of security breaches.

First, we give the most general definition of encoding in classical secret sharing schemes.

*Definition 12:* Let $\{1, \dots, n\}$ be $n$ participants or equivalently a set of $n$ shares. Let $X_1, \dots, X_n$ be random variables of $n$ shares. For each $i = 1, \dots, n$, $X_i$ denotes a share distributed to the $i$-th participant. Let $S$ be a random variable of a classical secret. Encoding of a classical secret sharing scheme is defined by a conditional probability $P(X_1, \dots, X_n|S)$ of $X_1, \dots, X_n$ given $S$.

In classical secret sharing, a dealer can realize advance sharing by generating the rest of shares based on given secret and shares already distributed. Thus the dealer needs to keep the information about shares already distributed until all shares are generated. Shares distributed in advance can leak out from the dealer. If shares distributed in advance leak out, forbidden sets are narrower than designed. It implies that smaller leakage allows an adversary to gain the partial information about a secret than designed and thus advance sharing of classical secret sharing can cause extra danger of security breaches. The dealer wishes to avoid extra danger of security breaches. To avoid extra danger of security breaches, the dealer would like not to keep any information about shares already distributed. What happens if the dealer does not keep any information about shares already distributed? Theorem 13 provides an answer to this question.

*Theorem 13:* Let $B$ be an advance-shareable set of shares which implies $I(S; \{X_i : i \in B\}) = 0$, where $I(\cdot;\cdot)$ is the mutual information [38]. Suppose that the dealer does not keep any information about a set of shares $B$ distributed before a given secret or equivalently $\{X_i : i \in B\} \leftrightarrow S \leftrightarrow \{X_i : i \in \overline{B}\}$ forms a Markov chain [38] (i.e., the conditional distribution of $\{X_i : i \in \overline{B}\}$ depends only on $S$ and is conditionally independent of $\{X_i : i \in B\}$ for a given $S$). Then the amount of information about a secret is determined by only a set of shares $\overline{B}$ or equivalently for any set of shares $D \subset B$ and $E \subset \overline{B}$, $I(\{X_i : i \in D \cup E\}; S) = I(\{X_i : i \in E\}; S)$.

*Proof:* Since $\{X_i : i \in B\} \leftrightarrow S \leftrightarrow \{X_i : i \in \overline{B}\}$ forms a Markov chain, $\{X_i : i \in D\} \leftrightarrow S \leftrightarrow \{X_i : i \in E\}$ forms a Markov chain. We have

$$\begin{aligned} & I(\{X_i : i \in D \cup E\}; S) \\ &= I(\{X_i : i \in E\}; S | \{X_i : i \in D\}) \\ &\quad + I(\{X_i : i \in D\}; S | \{X_i : i \in E\}) && (33) \\ &\le I(\{X_i : i \in E\}; S | \{X_i : i \in D\}) \\ &\quad + I(\{X_i : i \in D\}; S) && (34) \\ &\le I(\{X_i : i \in E\}; S | \{X_i : i \in D\}) + 0 && (35) \\ &\le I(\{X_i : i \in E\}; S), && (36) \end{aligned}$$

where (33) follows from chain rule for the mutual information [38, Theorem 2.5.2], (34) and (36) follow from a Markov chain $\{X_i : i \in D\} \leftrightarrow S \leftrightarrow \{X_i : i \in E\}$ and [38, Theorem 2.8.1], and (35) follows from the assumption $I(S; \{X_i : i \in B\}) = 0$. Since $I(\{X_i : i \in D \cup E\}; S) \ge I(\{X_i : i \in E\}; S)$, we have $I(\{X_i : i \in D \cup E\}; S) = I(\{X_i : i \in E\}; S)$. ■

Theorem 13 provides an answer to the question as follows: Suppose that the dealer does not keep any information about a set of shares $B$ already distributed. Then even if participants in $B$ dispose their shares distributed before a secret is given, the amount of information about a secret gained from shares of participants is the same. That is, distributing a set of shares $B$ is completely wasteful and advance sharing is worthless. Therefore, the dealer cannot avoid extra danger of security breaches described above if the dealer uses classical secret sharing.

In contrast, the rest of shares are generated only based on a given secret in our proposed quantum secret sharing scheme. Thus the dealer does not need to keep the information about shares already distributed. For this reason, even if security breaches expose dealer's storage before a secret is given, our proposed secret sharing is as secure as designed. Therefore, our proposed secret sharing has no extra danger of security breaches which is unavoidable in any classical secret sharing.

## VI. GILBERT-VARSHAMOV-TYPE SUFFICIENT CONDITION

In this section, we give a Gilbert-Varshamov-type sufficient condition for existence of our proposed secret sharing and then discuss its engineering implication.

*Theorem 14:* If positive integers $n, k, s, \delta_q, \delta_f, \delta_t$ with $\delta_f \ge \delta_t$ satisfy

$$\begin{aligned} & \frac{q^{n+k+s} - q^{n+s}}{q^{2n} - 1} \sum_{i=1}^{\delta_q - 1} \binom{n}{i} (q^2 - 1)^i \\ & + \frac{q^n - q^{n-s}}{q^{2n} - 1} \sum_{i=1}^{\delta_t - 1} \binom{n}{i} (q^2 - 1)^i \\ & + \frac{q^{n-s} - q^{n-k-s}}{q^{2n} - 1} \sum_{i=1}^{\delta_f - 1} \binom{n}{i} (q^2 - 1)^i < 1, \qquad (37) \end{aligned}$$

then there exist $C_S \subset C_R \subset C_{\max} = C_{\max}^{\perp s} \subset C_R^{\perp s} \subset C_S^{\perp s} \subset \mathbb{F}_q^{2n}$ such that $\dim C_S = n - k - s$, $\dim C_R = n - s$, $d_s(C_S^{\perp s}, C_R^{\perp s}) \ge \delta_q$, $d_s(C_R, C_S) \ge \delta_f$ and $d_s(C_{\max}, C_S) \ge \delta_t$.

*Remark 15:* By Theorem 1, a sufficient condition on forbidden sets remains the same as the conventional quantum secret sharing scheme [22]. Thus by [22, Theorem 6], the condition $d_s(C_R, C_S) \ge \delta_f$ implies that a set of $\delta_f - 1$ shares is forbidden. By Theorem 8, the condition $d_s(C_{\max}, C_S) \ge \delta_t$ implies that a set of $\delta_t - 1$ shares is advance-shareable.

*Proof:* Let $A(k, s)$ be the set of triples of linear spaces $(U, V, W)$ such that $\dim U = n - k - s$, $\dim V = n - s$, $\dim W = n$ and $U \subset V \subset W = W^{\perp s} \subset V^{\perp s} \subset U^{\perp s} \subset \mathbb{F}_q^{2n}$. For $\mathbf{e} \in \mathbb{F}_q^{2n}$, define $B_{U^{\perp s}}(k, \mathbf{e}) = \{(U, V, W) \in A(k, s) : \mathbf{e} \in U^{\perp s} \setminus V^{\perp s}\}$, $B_W(k, \mathbf{e}) = \{(U, V, W) \in A(k, s) : \mathbf{e} \in W \setminus V\}$ and $B_V(k, \mathbf{e}) = \{(U, V, W) \in A(k, s) : \mathbf{e} \in V \setminus U\}$.

For nonzero $\mathbf{e}_1, \mathbf{e}_2 \in \mathbb{F}_q^{2n}$, we have $|B_{U^{\perp s}}(k, \mathbf{e}_1)| = |B_{U^{\perp s}}(k, \mathbf{e}_2)|$, $|B_W(k, \mathbf{e}_1)| = |B_W(k, \mathbf{e}_2)|$ and $|B_V(k, \mathbf{e}_1)| = |B_V(k, \mathbf{e}_2)|$ by the almost same argument as [22, Proof of Theorem 15].

For each $(U, V, W) \in A(k, s)$, the number of $\mathbf{e}$ such that $\mathbf{e} \in U^{\perp s} \setminus V^{\perp s}$ is $|U^{\perp s}| - |V^{\perp s}| = q^{n+k+s} - q^{n+s}$. The number of quadruples $(\mathbf{e}, U, V, W)$ such that $\mathbf{0} \neq \mathbf{e} \in U^{\perp s} \setminus V^{\perp s}$ is

$$\sum_{\mathbf{0} \neq \mathbf{e} \in \mathbb{F}_q^{2n}} |B_{U^{\perp s}}(k, \mathbf{e})| = |A(k, s)| \times (q^{n+k+s} - q^{n+s}),$$

which implies

$$\frac{|B_{U^{\perp s}}(k, \mathbf{e})|}{|A(k, s)|} = \frac{q^{n+s} - q^{n+k+s}}{q^{2n} - 1}. \qquad (38)$$

Similarly we have

$$\frac{|B_W(k, \mathbf{e})|}{|A(k, s)|} = \frac{q^n - q^{n-s}}{q^{2n} - 1}, \qquad (39)$$

$$\frac{|B_V(k, \mathbf{e})|}{|A(k, s)|} = \frac{q^{n-s} - q^{n-k-s}}{q^{2n} - 1}. \qquad (40)$$

If there exists $(U, V, W) \in A(k, s)$ such that $(U, V, W) \notin B_{U^{\perp s}}(k, \mathbf{e}_1)$, $(U, V, W) \notin B_W(k, \mathbf{e}_2)$ and $(U, V, W) \notin B_V(k, \mathbf{e}_3)$ for all $1 \le \mathrm{swt}(\mathbf{e}_1) \le \delta_q - 1$, $1 \le \mathrm{swt}(\mathbf{e}_2) \le \delta_t - 1$ and $1 \le \mathrm{swt}(\mathbf{e}_3) \le \delta_f - 1$, then there exists a triple of $(U, V, W)$ with the desired properties. The number of $\mathbf{e}$ such that $1 \le \mathrm{swt}(\mathbf{e}) \le \delta - 1$ is given by

$$\sum_{i=1}^{\delta - 1} \binom{n}{i} (q^2 - 1)^i. \qquad (41)$$

By combining (38), (39), (40) and (41), we see that (37) is a sufficient condition for ensuring the existence of $(U, V, W)$ required in Theorem 14. ■

We will derive Theorem 16 as an asymptotic form of Theorem 14.

*Theorem 16:* Let $R \le 1$, $S \le 1$, $\varepsilon_t \le \varepsilon_f < 0.5$, $\varepsilon_q < 0.5$ be nonnegative real numbers. Define $h_q(x) = -\log_q x - (1 - x)\log_q(1 - x)$. For sufficiently large $n$, if

$$\begin{aligned} h_q(\varepsilon_q) + \varepsilon_q \log_q(q^2 - 1) &< 1 - R - S, \\ h_q(\varepsilon_t) + \varepsilon_t \log_q(q^2 - 1) &< 1 \text{ and} \\ h_q(\varepsilon_f) + \varepsilon_f \log_q(q^2 - 1) &< 1 + S, \end{aligned}$$

then there exist $C_S \subset C_R \subset C_{max} = C_{max}^{\perp s} \subset C_R^{\perp s} \subset C_S^{\perp s} \subset \mathbb{F}_q^{2n}$ such that $\dim C_S = n - \lfloor n(R+S) \rfloor$, $\dim C_R = n - \lfloor nS \rfloor$, $d_s(C_S^{\perp s}, C_R^{\perp s}) \geq \lfloor n\varepsilon_q \rfloor$, $d_s(C_R, C_S) \geq \lfloor n\varepsilon_f \rfloor$ and $d_s(C_{max}, C_S) \geq \lfloor n\varepsilon_t \rfloor$.

*Proof:* Proof can be done by almost the same argument as [22, Theorem 16]. ■

For sufficiently large $n$, the size of advance-shareable sets (i.e., $\lfloor n\varepsilon_t \rfloor - 1$ in Theorem 16) depends on only a prime power $q$ and independent of the size of classical secrets (i.e., $R$ in Theorem 16). We compare our proposed quantum secret sharing scheme with classical secret sharing schemes and then can make the following observation: Consider a scheme to share a $n \log_2 q$-bit classical secret to sufficient large number $n$ of participants by distributing a $\log_2 q$-bit or $\log_2 q$-qubit (i.e., $q$-qudit) share to each participant. When a dealer distributes a $\log_2 q$-bit share to each participant, all of each share depend on the secret and there are no advance-shareable shares. On the other hand, our proposed quantum secret sharing scheme distributing a $\log_2 q$-qubit share to each participants can make a set of $\lfloor n\varepsilon_t \rfloor - 1$ shares advance-shareable for $\varepsilon_t$ satisfying that $h_q(\varepsilon_t) + \varepsilon_t \log_q(q^2 - 1) < 1$. Particularly when $q = 2$, Theorem 16 implies that a set of roughly 19% of shares is advance-shareable independently of the size of classical secrets (i.e., $R$ in Theorem 16), as $h_2(0.19) + 0.19 \log_2 3 \simeq 1$ for sufficiently large $n$.

For sufficiently large $n$, Theorem 16 gives a sufficient condition on existence of our proposed secret sharing schemes with a set of $\lfloor n\varepsilon_f \rfloor - 1$ shares being forbidden and a set of $\lfloor n\varepsilon_t \rfloor - 1$ shares being advance-shareable, which is independent of the size of classical secrets (i.e., $R$ in Theorem 16).

## VII. CONCLUSION

In our paper, we propose quantum secret sharing with the capabilities in designing of access structures more flexibly and realizing higher efficiency beyond those of classical secret sharing, that can distribute some shares before a given secret. We clarify a necessary and sufficient condition on advance-shareable sets in Sect. III. Our proposed quantum secret sharing can make the size of forbidden sets and advance-shareable sets larger than classical linear secret sharing scheme, which is demonstrated in Sect. IV by comparing our proposed quantum secret sharing scheme from the Reed–Solomon codes with a ramp version of Shamir's scheme. In Sect. V, we prove that in classical secret sharing, a dealer needs to keep the partial information about shares already distributed until all shares are generated, which causes extra danger of security breaches. On the other hand, the dealer does not need to keep the information about shares already distributed in our proposed quantum secret sharing. Thus our proposed quantum secret sharing has no extra danger of security breaches. In Sect. VI, we give a sufficient condition on existence of our proposed secret sharing schemes with the given size of forbidden and advance-shareable sets, which is independent of the size of classical secrets. Therefore, our proposal can provide a useful method of advance sharing when a dealer unable to communicate with some participants after the dealer obtains a secret.

## ACKNOWLEDGMENT

The authors would like to thank Prof. Tomohiko Uyematsu for a helpful advice and Dr. Seok Hyung Lie for informing them of the related work [14]. An earlier version of this paper was presented in part at QCrypt 2022, Taipei City, Taiwan, August 29–September 2, 2022.

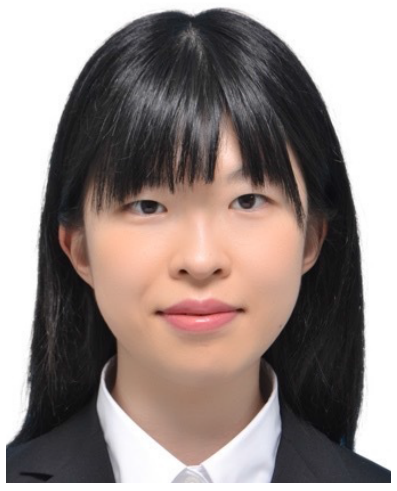

**RINA MIYAJIMA** was born in Yokohama, Japan, in May 1999. She received the B.Sc. degree in mathematics from the Tokyo Institute of Technology, Japan, in 2022, where she is currently pursuing the M.E. degree with the Department of Information and Communications Engineering.

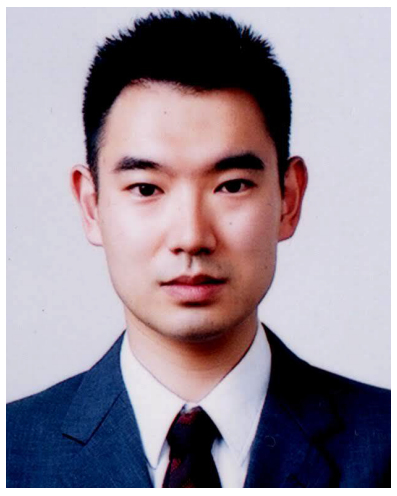

**RYUTAROH MATSUMOTO** (Member, IEEE) was born in Nagoya, Japan, in November 1973. He received the B.E. degree in computer science, the M.E. degree in information processing, and the Ph.D. degree in electrical and electronic engineering from the Tokyo Institute of Technology, Tokyo, Japan, in 1996, 1998, and 2001, respectively.

From 2001 to 2004, he was an Assistant Professor with the Department of Information and Communications Engineering, Tokyo Institute of Technology, where he was an Associate Professor, from 2004 to 2017. From 2017 to 2020, he was an Associate Professor with the Department of Information and Communication Engineering, Nagoya University, Nagoya, Japan. Since 2020, he has been an Associate Professor, and promoted to a Full Professor, in August 2022, with the Department of Information and Communications Engineering, Tokyo Institute of Technology. In 2011 and 2014, he was as a Velux Visiting Professor with the Department of Mathematical Sciences, Aalborg University, Aalborg, Denmark. His research interests include error-correcting codes, quantum information theory, information theoretic security, and communication theory.

Dr. Matsumoto was a recipient of the Young Engineer Award from IEICE and the Ericsson Young Scientist Award from Ericsson Japan in 2001. He was also a recipient of the Best Paper Awards from IEICE in 2001, 2008, 2011, and 2014.